# Reconfigurable optical computing based on graphene

Abolfazl Ghanbari, Ali Mahjoori*

*al.mahjoori@gmail.com

Department of Electrical and Computer Engineering, University of Sistan and Baluchestan, Iran, 98167-45845

**Optical computing has recently attracted a great deal of interest as it offers the ability to process data in a parallel manner. In this report, an optical computing system based on a metamaterial structure made of graphene is designed and demonstrated. It is shown that the proposed structure is able to do different operations such as taking derivative, and integration of an incident field. In addition to this, it is shown that the proposed structure is reconfigurable due to the possibility of tuning the surface conductivities of the graphene flakes. The proposed configuration not only goes beyond the major restriction of traditional electronic computers, but also provides a dynamic operation.**

## 1. INTRODUCTION

Traditional electronic computers were restricted by several issues, including their bulky configuration and slow speed. The ability to process data parallelly has allowed optical computation [1-8] to overcome these restrictions, enabling various applications such as equation solving [9], imaging and edge detection [10-16], and topological data processing [17,18].

The concept of computational metamaterials was proposed for the first time in [19], performing different types of mathematical operations like convolution, integration, etc. This was done by manipulating the incident signals as they were propagating in a metamaterial slab, such that the wave profile at the output of the slab becomes proportional to the desired one.

In another area of research, recently, there has been a great interest in graphene-based metamaterials and structures, due to their capability in proving reconfigurability. As a matter of fact, over the last 20 years, the interesting conductive features of graphene has attracted a lot of attention in applications such as sensing, switching, modulators, absorbers, waveguides, antennas, etc [20-43].

Motivated by these findings, this Letter is aimed to prove the importance of graphene for doing reconfigurable optical computing. As a matter of fact, it is shown that a metamaterial built from graphene elements can be designed to perform different operation functions. Moreover, it is shown that such operation can be manipulated in a desired manner by changing the chemical potential of the graphene flakes. The proposed graphene-based structure not only circumvents the major potential drawbacks of traditional computers, but also offers a dynamic functionality.

The paper is organized as follows. In Section II, the optical model of graphene is discussed. In section III, the proposed structure, performing different operation is introduced. In section III, the performance of the proposed device is investigated. Finally, in section IIII, a conclusion of our findings is provided.

## 2. MODEL OF GRAPHENE

Graphene is a material composed of carbon atoms arranged in a hexagonal type of lattice. In fact, graphene is known to be the 2D version of graphite. Usually, the optical properties of graphene are modelled using a formula known as Kubo formula, expressing its surface conductivity as

$$\sigma = \frac{e^2 K_B T}{\pi \hbar^2} \frac{i}{\omega + i \tau^{-1}} [\frac{E_F}{K_B T} + 2\ln(e^{\frac{-E_F}{K_B T}} + 1)] + \frac{ie^2}{4\pi\hbar} \ln[\frac{2E_F - \hbar(\omega + i \tau^{-1})}{2E_F + \hbar(\omega + i \tau^{-1})}] + \frac{ie^2}{4\pi\hbar} \ln[\frac{2E_F - \hbar(\omega + i \tau^{-1})}{2E_F + \hbar(\omega + i \tau^{-1})}]$$

(1)

in which e is the charge of electron, $K_B$ is Boltzmann constant, $T$ is the temperature in Kelvin, $E_f$ is the Fermi level, $\omega$ is the angular frequency, $h$ is Plank constant, and $\tau$ is relaxation time. In the following, the model expressed in Eq. 1 is used to model the optical properties of graphene.

## 3. PROPOSED GRAPHENE-BASED METASTRUCTURE

The geometry of the proposed graphene-based structure, that is aimed to perform different kind of operation is shown in Figure. 1a. The structure composes of a graphene array of square thin plates placed on top of a substrate, namely silicon dioxide. It is worth mentioning that the size of the flakes is very small than the wavelength. By changing the

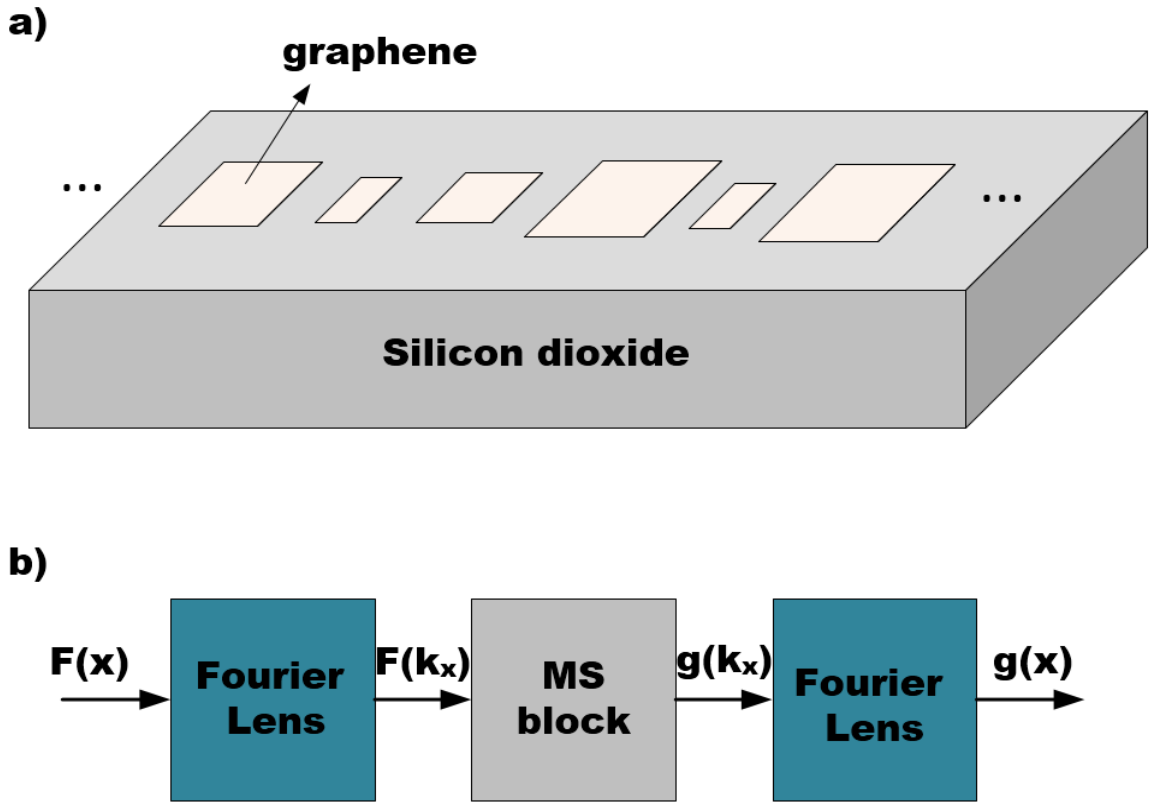

**Figure1: a) Proposed graphene-based optical computing system. b) Block diagram sketch of the proposed graphene-based optical computing system.**

length and the width of the squares, it is possible to achieve almost $2\pi$ phase shift, while having a high level of transmission (the interested reader is referred to Supplementary materials for more information about this). It is also good to note that two graded index Fourier lens are also employed before and after the proposed structure to in order to Fourier transform the input and output fields from the proposed configuration. These Fourier lenses are important blocks in our proposed optical computing system, whose block diagram is shown in Figure. 1b. The proposed block diagram consists of two Fourier transform block, explained earlier, as well as the metasurface slab multiplying the transfer function of the desired operator to the Fourier transformed version of the incident field. In the following sections, we illustrate the possibility of performing different types of mathematical operations by simply varying the sizes of these graphene sheets.

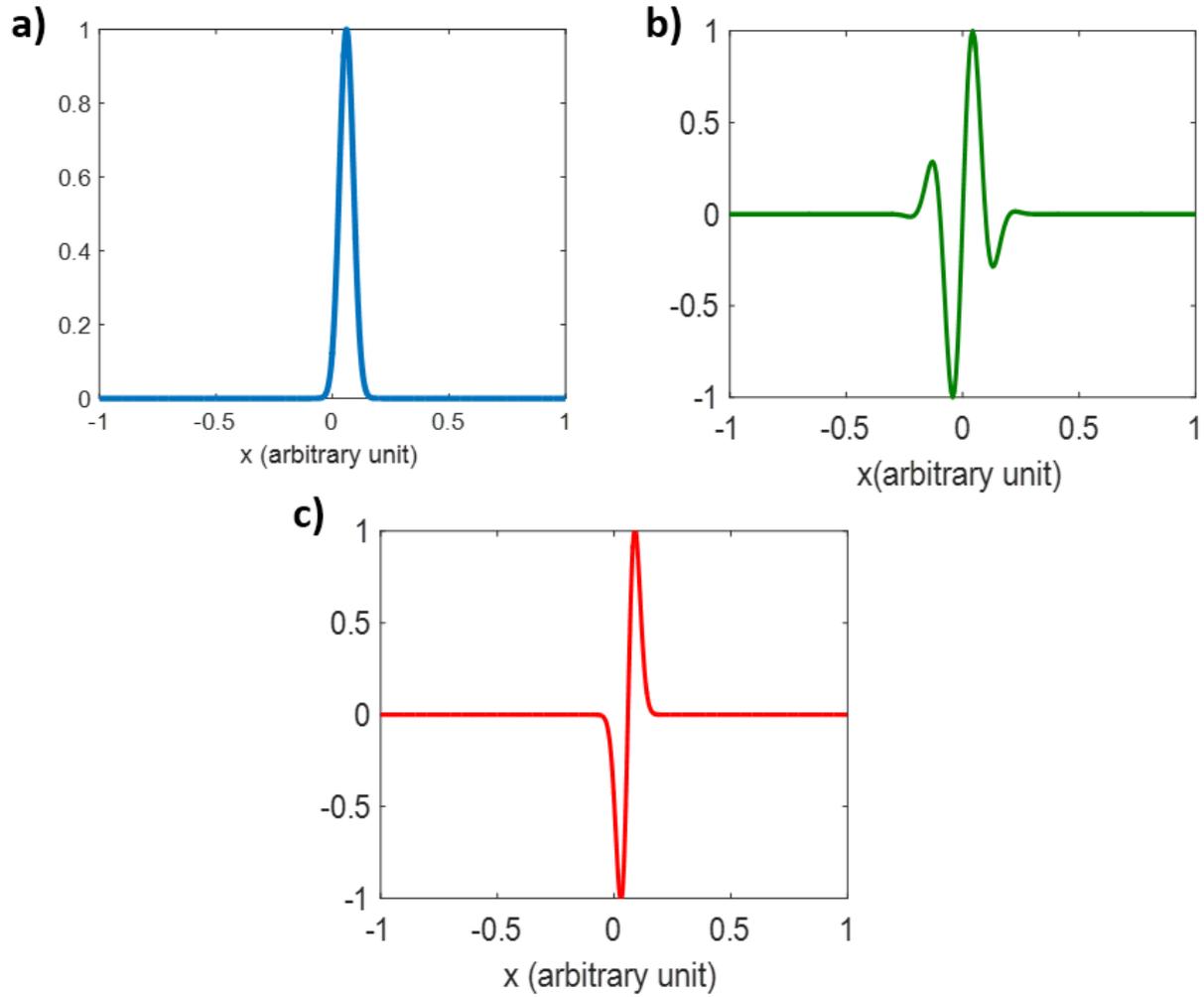

**Figure2: a) An incident Gaussian beam is impinging the configuration. b) Resulting output field from the metamaterial slab. c) Ideal derivative of the signal in panel a.**

## 4. PERFORMANCE ANALYSIS OF THE PROPOSED STRUCTURE

In this section, we verify the feasibility of calculating first order derivative of an optical field impinging to our proposed graphene-based structure. To this end, as we explained in the previous section, it is needed to spatially vary the size of the squares according to the corresponding desired transfer function $T_{r1} = ik_x$. The interested reader is referred to Table I of Supplementary materials to find the values of the dimensions, needed to realize the transfer function $T_{r1}$.

To evaluate the performance of such a differentiator, we consider a typical optical Gaussian beam, whose profile is shown in the inset of Fig. 2a. The corresponding profile of the output beam is shown in Fig 2b. Fig. 2c reports the exact shape of the first order derivative of the input signal. Comparing the latter two figures with each other, one induces that the proposed differentiator is functioning perfect.

Next, we analyse the possibility of calculating integration of an optical field impinging to our proposed graphene-based structure. To this end, as usual, we change the size of the squares according to the desired transfer function the integrator, i.e $T_{r2} = 1/ik_x$. The interested reader is referred to Table II of Supplementary materials to find the corresponding values of the dimensions, required to realize the transfer function $T_{r2}$. To verify the performance of the proposed integrator, the incident beam is supposed to be a Gaussian derivative one, whose profile is shown in the Fig. 3a. The calculated profile of the output beam is shown in Fig 3b. Fig. 3c depicts the exact shape of the integrated version of the input signal. Comparing Fig. 3b with Fig. 3c validates the good functionality of the proposed integrator.

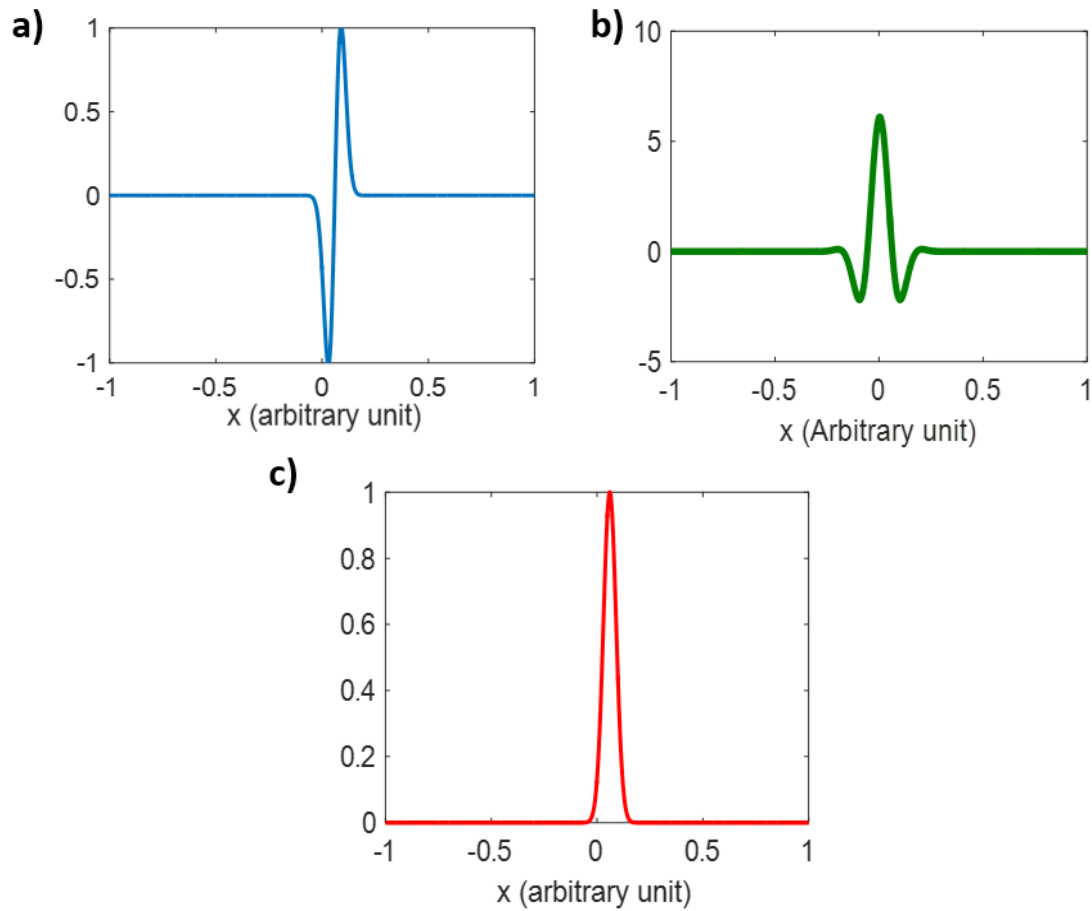

**Figure2: a) An incident Gaussian beam is impinging the configuration. b) Resulting output field from the metamaterial slab. c) Ideal derivative of the signal in panel a.**

Insert PSN Here

## 5. RECONFIGURABILTY OF THE PROPOSED GRAPHENE-BASED STRUCTURE

As the proposed metamaterial structure is composed of graphene, it is possible to dynamically tune it to different spectral ranges, without changing the geometry. To this end, one has to change the values of the chemical potentials. The latter can be done by applying an external bias voltage to the flakes and changing it. To demonstrate such a good behavior, we increase of operation of the aforementioned differentiator (by 10 percent) designed in the previous section and try to maintain the original functionality of the differentiator by changing the values of the chemical potentials (the reader is referred to Table III of the Supplemental materials for the exact values of the chemical potentials used here). The corresponding input and output beams are represented in Figures 4 a and b respectively. It is obvious that the proposed differentiator has been perfectly reconfigured to the desired frequency range.

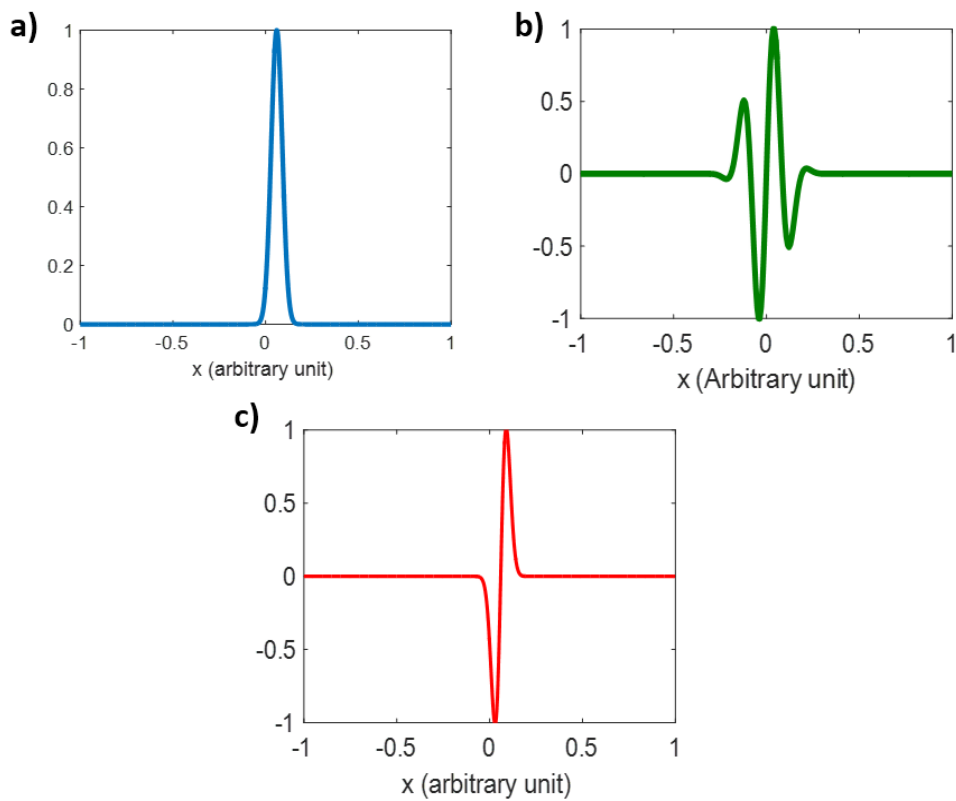

**Figure4: Reconfigurability of the proposed configuration. The figure repeats the result of Fig. 2 for an operation frequency of 1.1 that of that figure.**

## 6. CONCLUSION

As a conclusion, inspired by the recent advances in the fields of metamaterials and graphene, the importance of graphene for doing optical computing tasks such as differentiation and integration was demonstrated. By proposing a metamaterial built from graphene elements of square shape, it was shown that it is possible to realize transfer functions of interest. Moreover, it was confirmed that the proposed graphene-based system can be manipulated in a desired manner by changing the chemical potential of the graphene flakes.